\documentclass[12pt]{spieman}  % 12pt font required by SPIE;
\usepackage{amsmath,amsfonts,amssymb}
\usepackage{graphicx}
\usepackage{setspace}
\usepackage{tocloft}
\usepackage{array}
\usepackage{verbatim}
\usepackage{subcaption}
\usepackage{fontenc}
\usepackage{fancyhdr}

\newcolumntype{P}[1]{>{\centering\arraybackslash}p{#1}}
\newcommand{\nuvu}{N\"uv\"u } 
\usepackage{lineno}
\usepackage{soul}

\usepackage[headsep=1cm]{geometry}
\fancypagestyle{firstpage}
{
    \fancyhead[L]{{\copyright} 2019. All rights reserved.}
    \fancyhead[R]{}
}

\title{Delta-doped Electron Multiplying CCDs for FIREBall-2.}
\author[a,*]{Gillian Kyne}
\author[b]{Erika T. Hamden}
\author[a]{Shouleh Nikzad}
\author[c]{Keri Hoadley}
\author[a]{April Jewell}
\author[a]{Todd Jones}
\author[a]{Michael Hoenk}
\author[a]{Samuel Cheng}
\author[c]{D.~Christopher Martin}
\author[c]{Nicole Lingner}
\author[d]{David Schiminovich}
\author[e]{Bruno Milliard}
\author[e]{Robert Grange}
\author[f]{Olivier Daigle}
\affil[a]{Jet Propulsion Laboratory, California Institute of Technology, 4800 Oak Grove Drive, Pasadena, CA 91109, USA}
\affil[b]{University of Arizona and Steward Observatory, 933 N Cherry Ave, Tucson, AZ 85721, USA}
\affil[c]{California Institute of Technology, Division of Physics, Math, and Astronomy, 1200 E California Blvd, MC 278-17, Pasadena, CA 91105, USA}
\affil[d]{Department of Astronomy, Columbia University, 550 W. 120th Street, New York, NY 10027, USA}
\affil[e]{Laboratoire d'Astrophysique de Marseille, 38 Rue Fr\'{e}d\'{e}ric Joliot Curie, 13013 Marseille, France}
\affil[f]{\nuvu Cameras, 603-355 Peel, Montr\'{e}al, QC, H3C 2G9, Canada}

\cftpagenumbersoff{figure}
\cftpagenumbersoff{table} 
\begin{document}
%\linenumbers
%\copyrighttext{{\copyright} 2019. All rights reserved.}
%{{\copyright} 2019. All rights reserved.}

\thispagestyle{firstpage}
\maketitle

\begin{abstract}
We present the status of on-going detector development efforts for our joint NASA/CNES balloon-borne UV multi-object spectrograph, the Faint Intergalactic Redshifted Emission Balloon (FIREBall-2; FB-2). FB-2 demonstrates a new UV detector technology, the delta-doped Electron Multiplying CCD (EMCCD), in a low risk suborbital environment, to prove the performance of EMCCDs for future space missions and Technology Readiness Level (TRL) advancement. EMCCDs can be used in photon counting (PC) mode to achieve extremely low readout noise ($<$1 electron). Our testing has focused on reducing clock-induced-charge (CIC) through wave shaping and well depth optimization with a \nuvu V2 CCCP Controller, measuring CIC at 0.001 e$^{-}$/pixel/frame. This optimization also includes methods for reducing dark current, via cooling, and substrate voltage levels. We discuss the challenges of removing cosmic rays, which are also amplified by these detectors, as well as a data reduction pipeline designed for our noise measurement objectives. FB-2 flew in 2018, providing the first time an EMCCD was used for UV observations in the stratosphere. FB-2 is currently being built up to fly again in 2020, and improvements are being made to the EMCCD to continue optimizing its performance for better noise control.
\end{abstract}


% Include email contact information for corresponding author
{\noindent \footnotesize\textbf{*}Gillian Kyne,  \linkable{gillian.kyne@jpl.nasa.gov} }

\begin{spacing}{2}   % use double spacing for rest of manuscript

%%%%%%%%%%%%%%%%%%%%%%%%%%%%%%%%%%%%%%%%%%%%%%%%%%%%%%%%%%%%%
\section{Introduction}
\label{sec:intro}  % \label{} allows reference to this section

FIREBall-2 is a balloon-borne ultraviolet (UV) multi-object spectrograph. It was designed to discover and map faint emission from the circumgalactic medium (CGM) around low redshift galaxies (z $\sim$ 0.7). This mission is an international collaboration between a number of institutes in both the US (Caltech, JPL, Columbia University) and France (Centre National d'\'Etudes Spatiales (CNES) and Laboratoire d'Astrophysique de Marseille (LAM)). The instrument and telescope is an upgraded version of FIREBall-1, which flew twice, in 2006\cite{2008Tuttle} and 2009\cite{2010Milliard}. The spectrograph has been redesigned to increase the field of view, throughput, and number of observed targets\cite{2014Grange,2016Grange}, yielding a factor of 30 increase in overall sensitivity. This should also provide multiple detections of the CGM in emission for the first time at UV wavelengths. 

\begin{figure}%[h]%{l}{0.75\textwidth}
   % \vspace{-0.3cm}
    \centering
    \caption{Instrument Signal-to-Noise/Sensitivity Calculation comparing FB-1, FB-2 (2018), \& FB-2 (2020$+$). This figure was taken from the most recently awarded APRA proposal summarizing past and future performance.}
    \includegraphics[width=\textwidth]{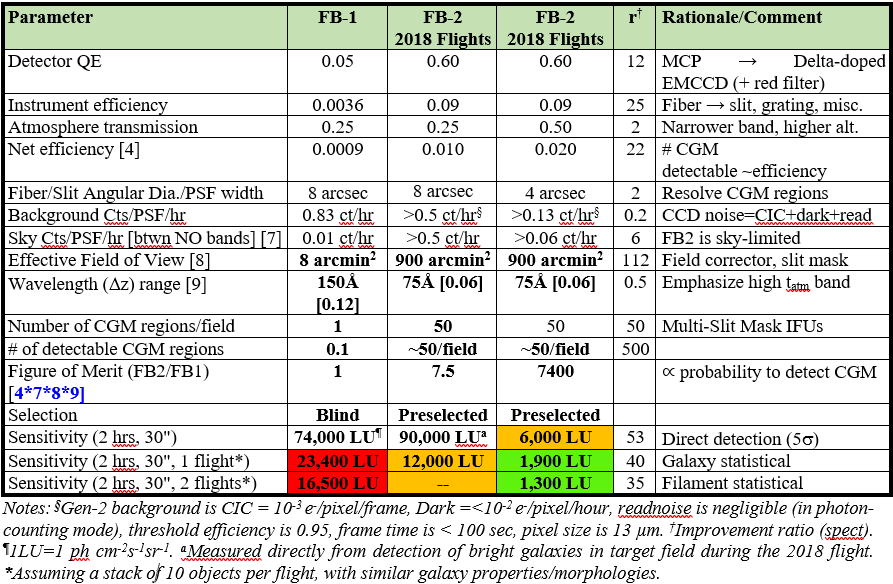}
    \label{fig:snr}
    \vspace{-0.8cm}
\end{figure}

FIREBall-2 is an excellent platform for advancing the TRL of new technologies. A fundamental requirement for achieving our science is improving our instrument sensitivity; our biggest component-level improvement has come from a high-efficiency, low-noise photon counting UV CCD. This technology is responsible for a factor of 13 increase in quantum efficiency (QE) on FIREBall-2 when compared to the FB-1 detector (Figure \ref{fig:snr}). This is a revolutionary change, and one that will be leveraged by NASA for future missions. When identifying essential UV technologies of the future, the AURA report (Association of Universities for Research in Astronomy) describes the need for a detector which can achieve the 'triple crown...high quantum efficiency, low read noise/photon counting, and low dark current.' This detector technology will meet this challenge.

The FIREBall detector team have collaborated with \nuvu Cameras who have developed custom controller hardware to readout these devices. The CCD201-20s have a 1k$\times$1k active area and a 1k$\times$1k storage area. They are usually read out in frame transfer (FT) mode, however FB-2 uses the full 2k$\times$1k in line transfer mode. This makes it possible for a larger field-of-view (FOV). 
\par
\nuvu are also developing the flight controller/electronics for the Wide Field Infrared Survey Telescope (WFIRST) coronograph instrument \cite{Harding2016}. For the past 6 years FB-2 has used a \nuvu V2 controller for laboratory characterization and testing before its integration with the spectrograph in 2016. FIREBall-2020 will be upgraded to using a \nuvu V3 controller, which can be adapted for radiative cooling, something which was not possible with the V2. Testing with the V3 controller is already underway and subsequent FIREBall missions will complement TRL advancement for this hardware.

%CNES is providing the spectrograph, gondola, and gondola flight support team. 
FIREBall-2 launched for the first time on September 22$^{nd}$, 2018 from the Columbia Scientific Balloon Facility (CSBF) in Fort Sumner, New Mexico. The team demonstrated a successful launch and operation of the redesigned spectrograph and in particular, the first delta-doped Electron Multiplying CCD (EMCCD) used for science in a space-like environment.

%Mission enabling technology: NEXUS detectors
\par

%EMCCDs were chosen for this mission due to their photon counting (PC) abilities; a mode used to achieve extremely low readout noise (< 1 e$^{-}$). This detector has been processed at JPL to provide UV sensitivity using a delta-doping procedure, which is discussed further in \cite{6,8}. The process yields a device with 100\% internal quantum efficiency, limited only by surface reflections. Anti-reflection coatings to minimize this reflectance bring the quantum efficiency of the FIREBall-2 device up to > 65\% from 200-210 nm. FIREBall-2 uses an e2V CCD201-20 in photon counting mode. This device is normally operated in frame transfer mode, however, the FIREBall-2 instrument makes use of the entire 1K$\times$2K array.

\section{Electron Multiplying CCDs}
EMCCDs \cite{Mackay2001,Jerram2001} are a developing technology becoming more relevant for use in astrophysics applications, with a number of upcoming Flagship-class missions/concepts (WFIRST, LUVOIR, HabEx) and sounding rocket (SHIELDS -- Spatial Heterodyne Interferometric Emission Line Diagnostic Spectrometer), taking advantage of the new technology for unique science applications. These devices have been developed primarily for ``photon counting'' capabilities, which renders the read noise to $\sim$1 electron (e). EMCCDs operate similarly to a nominal CCD but have a second serial register containing additional pixels, where the second serial register clock (R2) is replaced with a high voltage (HV) clock. As individual electrons pass through the multiplication register, they are multiplied via impact ionization, see Jerram \cite{Jerram2001}. The exact multiplication gain achieved is a stochastic process controlled by the maximum voltage of the HV clock. Each individual transfer has a relatively low probability of multiplication ($\sim$2\%), but when an electron passes through all 604 multiplication pixels (as is for the case of the Teledyne-e2V CCD201-20, which are used in this study), the final number of electrons generated at the end pixel ends up being large \cite{Daigle2008,Daigle2010,Daigle2014a}. The testing described in this work uses an EMgain of $\sim$1000 e$^{-}$/e$^{-}$, or 1000 electrons created for a photon event registered in the image area. The exact value multiplication gain is not measured due to the random nature of this process, only an average value over a large number of pixels or frames.

\begin{figure}[hb]
\includegraphics[width=1.0\linewidth]{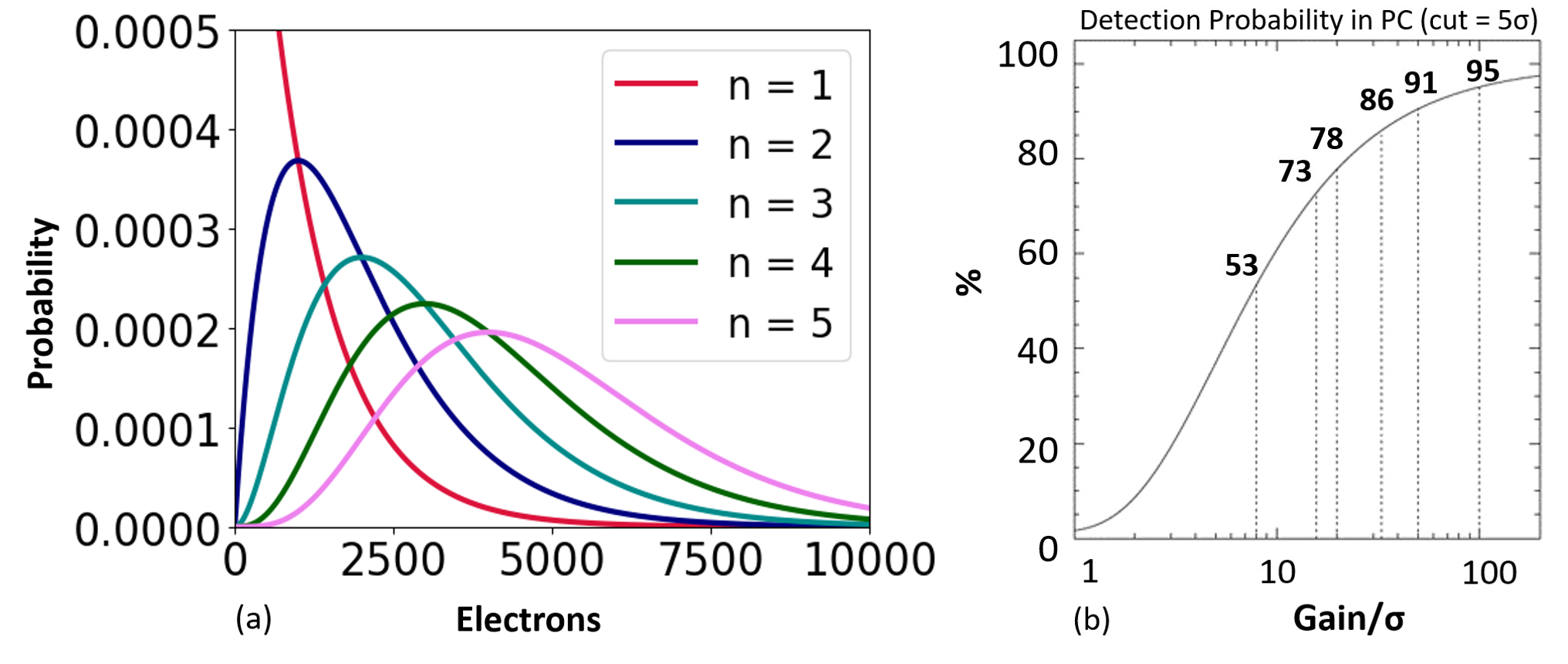}
\caption{Noise losses in EMCCDs: The multiplication process is stochastic in nature, it is only possible to measure the average multiplication gain from data. When more than 1 electron is incident on the detector, an uncertainty is created in how many electrons were at the outset. This can be seen from the overlap in Figure \ref{fig:emccds_noise_sources} (a). This plot was generated from Basden \cite{Basden2003}, Appendix A5. The noise from n $>$ 1 is called an Excess Noise Factor (ENF), which adds a $\sqrt{2}$ at high EMgain. Threshold losses occur when the multiplication gain is not sufficient to count $>$95\% of incident photons/electrons above the read noise, as seen in Figure \ref{fig:emccds_noise_sources} (b).}
\label{fig:emccds_noise_sources}
\end{figure}

The major advantage of using EMgain is that it increases the signal from a single photoelectron to a value much larger than the on-chip amplifier read noise. This allows single photon events to be detected by a simple thresholding process\cite{Tulloch2011,Harpse2012}. In general, pixels with counts greater than about $\sim$5 times the read noise (5-$\sigma$) are considered to have had 1 photoelectron event, and pixels below this threshold have zero events. However, there are some complications as a result of this process. Since the multiplication process is stochastic, not all photoelectrons detected will be amplified above this 5-$\sigma$ threshold. The ratio of EMgain to read noise provides an estimate for the number of events which are not counted because of low amplification\cite{Daigle2010} Figure \ref{fig:emccds_noise_sources} (b). Daigle \cite{Daigle2008} shows how this is calculated in Equations 1 and 2. 

Additional sources of noise come from clock-induced-charge (CIC) and dark current. A standard CCD cooled below -85$^{\circ}$C has thermal (dark current) and CIC noise sources present, but both are below the typical read noise (assuming a few electrons) and are thus undetectable. However, an EMCCD amplifies dark current and CIC noise in the same way as photoelectrons. Dark current is reduced by operating in inverted mode and/or by cooling the device. As we need to also minimize CIC, choosing to operate in non-inverted mode operation (NIMO -- low voltage substrate (VSS), 0V) achieves this allowing us to explore lower temperatures (10 - 25$^{\circ}$C lower) to reduce dark noise. There is evidence for a lower limit to dark rate where this rate plateaus at temperatures below -110$^{\circ}$C \cite{Tulloch2010PhDT,Daigle2012a,Kyne2016}. We are continuing this work with the \nuvu V3 controller but our preliminary results show that detector noise through poor charge transfer efficiency (CTE) increases, at very low temperatures, making the dark rate more difficult to measure. 
This temperature effect will be discussed further in Section \ref{data_redux}. It is worth noting that, depending on the science application, a lower event threshold can be applied to increase the number of statistical photoelectron detections.
\par
EMCCDs can also be used for non-photon counting applications - it is possible to operate them in a conventional mode, where the EMCCD is read out like normal CCD (without multiplication gain). An EMCCD can also be read out using a low value of multiplication gain, this is known as analog mode where the incoming signal is $>$ 1 e$^{-}$/pix/frame. In this mode the EMgain \footnote{From here on out the electron multiplication gain will be called EMgain.} used is only a few 100 e$^{-}$/e$^{-}$, which simply boosts the detected signal. In this mode, because the photon count rate is expected to be higher than 1 e$^{-}$/pix/frame, the image is subject to excess noise factor (ENF)\cite{Daigle2009}. This noise has a $\sqrt{2}$ value at high EMgain, see Figure \ref{fig:emccds_noise_sources}. This stochastic noise source, noticeable in low light conditions, affects the signal-to-noise ratio in the image, acting as if the QE of the sensor was halved. It becomes important that the multiplication gain is measured accurately, as this value is necessary to recover the original number of electrons produced by the source.
\par
Kyne \cite{Kyne2016} has presented results of dark current testing in both an engineering-grade and delta-doped EMCCDs\cite{Kyne2016}. In summary, low dark current and CIC noise is achieved through wave shaping, well-depth optimization (to reduce the CIC), and device cooling (to reduce dark current). For FIREBall-2, the optimization of these devices using a \nuvu V2 Controller has been performed using a dedicated test bed at Caltech. These devices are operated in NIMO to reduce CIC and cooled lower than the nominal -85$^{\circ}$C to lower the dark current rate. We adjust the controller clocking to reduce CIC through wave shaping and reduced image clock wells. Daigle\cite{Daigle2010} has reported that, by using sinusoidal- and triangular-shaped clocks, they can reduce CIC by a factor of 10 at the lowest readout frequencies\cite{Daigle2010}. They also found that clock-shaping does not cause a scaled increase in CIC with decreased pixel readout speed\cite{Daigle2010}. Currently, testing continues to find an optimized clocking scheme (using the \nuvu V3 controller) to achieve both low dark current and CIC, without a decrease in CTE and, hence, reduced device QE/throughput.

\subsection{Delta-doped EMCCDs}
FIREBall-2 utilizes an upgraded camera system, which includes an EMCCD (a T-e2V CCD201-20 device) optimized for the FB-2 bandpass (190 - 215 nm), which is achieved through a novel 2D-doping and anti-reflection (AR) coating process developed at the Microdevices Laboratory at JPL\cite{Nikzad2011,Hamden2015,Nikzad2016,Nikzad2017}. These combined technologies improve the FB-2 instrument performance by more than an order of magnitude\cite{Hamden2015}. The 2D-doping processes (also called delta-doping or superlattice-doping) has been successfully applied to a variety of silicon detector architectures\cite{Hoenk2014,Nikzad2016,Nikzad2017}. Briefly, the end-to-end post-fabrication, back-illumination process comprises a wafer-scale bonding, backside thinning to the photosensitive epitaxial silicon layer, followed by 2D doping, and ends with AR/filter coatings. The 2D-doping step results in stable device response at or near the reflection limit ($\sim$100\% internal QE) for wavelengths spanning soft X-rays to the near infrared\cite{Hoenk1992,Nikzad1994,Hoenk2013}. External QE can then be tailored and optimized with custom AR coatings and filters. %The 2D-doped UV EMCCD for FIREBall included a simple AR coating designed to maximize detector response within the FIREBall bandpass (190-215 nm). For a detailed discussion of the development of the AR coating for FIREBall-2 refer to our prior publications \cite{Nikzad2012,Hamden2014spie,Jewell2015,Hamden2016jatis}. 

\section{FIREBall-2 Laboratory \& Flight Hardware}
\label{sec:testing_setup}

FB-2 uses a \nuvu V2 CCD Controller for Counting Photons (CCCP)\cite{Daigle2008}, which was selected for its flexible waveform shaping ability and fast clocking speeds. The V2 controller - flown by FB-2 in September 2018 - has a readout rate of 10 MHz. \nuvu have since upgraded their electronics to a V3 controller that provides more readout options, including a 1 MHz high-voltage clock and greater resolution on clocking shapes. The \nuvu controller is connected to a custom Printed Circuit Board (PCB) with a SAMTEC EQCD high speed coax cable. 
\par
%A more detailed discussion of the laboratory set up at Caltech can be found in \cite{Kyne2016}. 

\begin{figure}[h]
\includegraphics[width=1.0\linewidth]{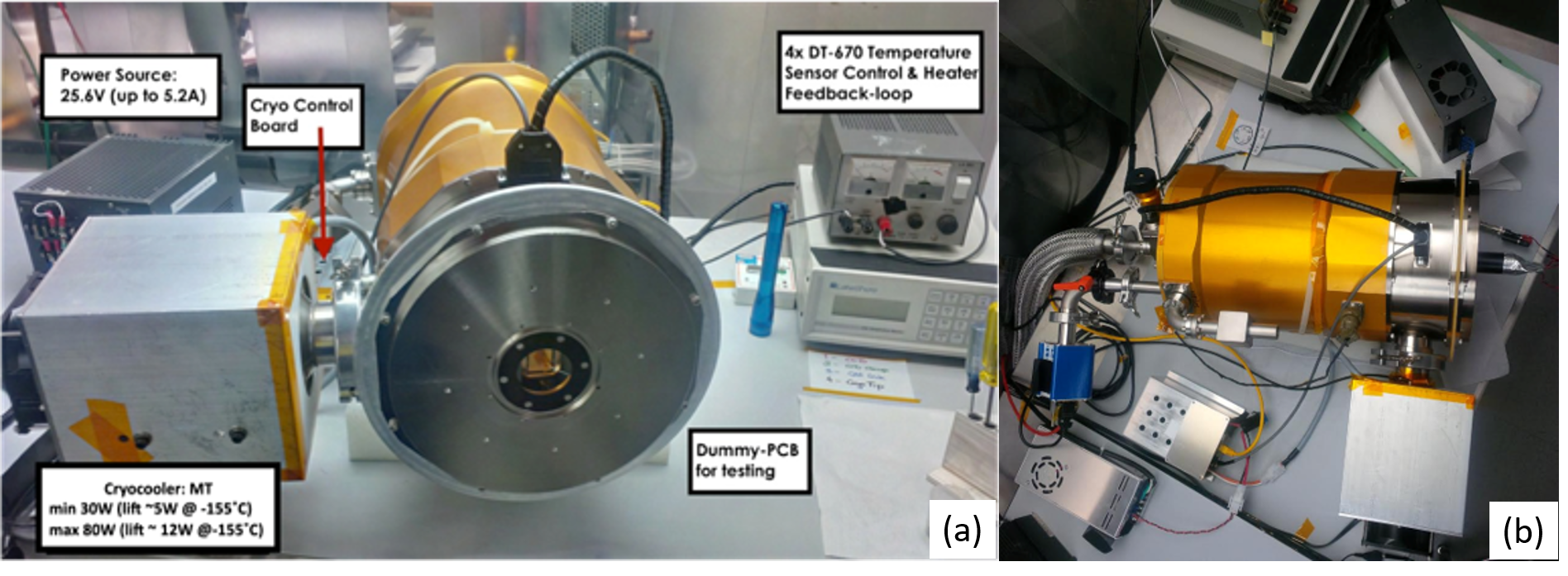} 
%\caption{Front-side view of the laboratory test setup. The cryocooler used for lab testing is used here to calibrate thermal sensors for the front and backside of a non-operational EMCCD. The Lakeshore unit used to read and control temperature is pictured on the RHS (top) where silicon diodes were used to measure temperature.}
\caption{Pictured is a front-side (a) and top view (b) of the laboratory setup at Caltech, first tested in 2015. The MT cryocooler from Sunpower was used for all laboratory testing but a CT model was used for flight as it has a higher cooling power range necessary for the flight hardware. The \nuvu controller pictured was the same one used for the 2018 flight.}
\label{fig:lab_setup}
\end{figure}

%Future setups have been based on this model.
Our PCB design was based on specifications provided with the \nuvu controller to achieve low read noise and provide flexibility in testing in the laboratory in preparation for flight. The PCB includes RC filters, located close to the sensor pins, to reduce residual noise ripples in the signal traces. A best practice of ground planes, including grounding paths, are used around the traces to give a low capacitance and inductance path for current at high frequencies. The ground planes help to further reduce noise in the transmission lines. The flight board was tested in the custom Caltech test bed, as pictured in Figure \ref{fig:lab_setup}, where thermal, vacuum, noise, and QE characterization was performed\cite{Kyne2016}.
\par
\begin{figure}[h]
\includegraphics[width=1.0\linewidth]{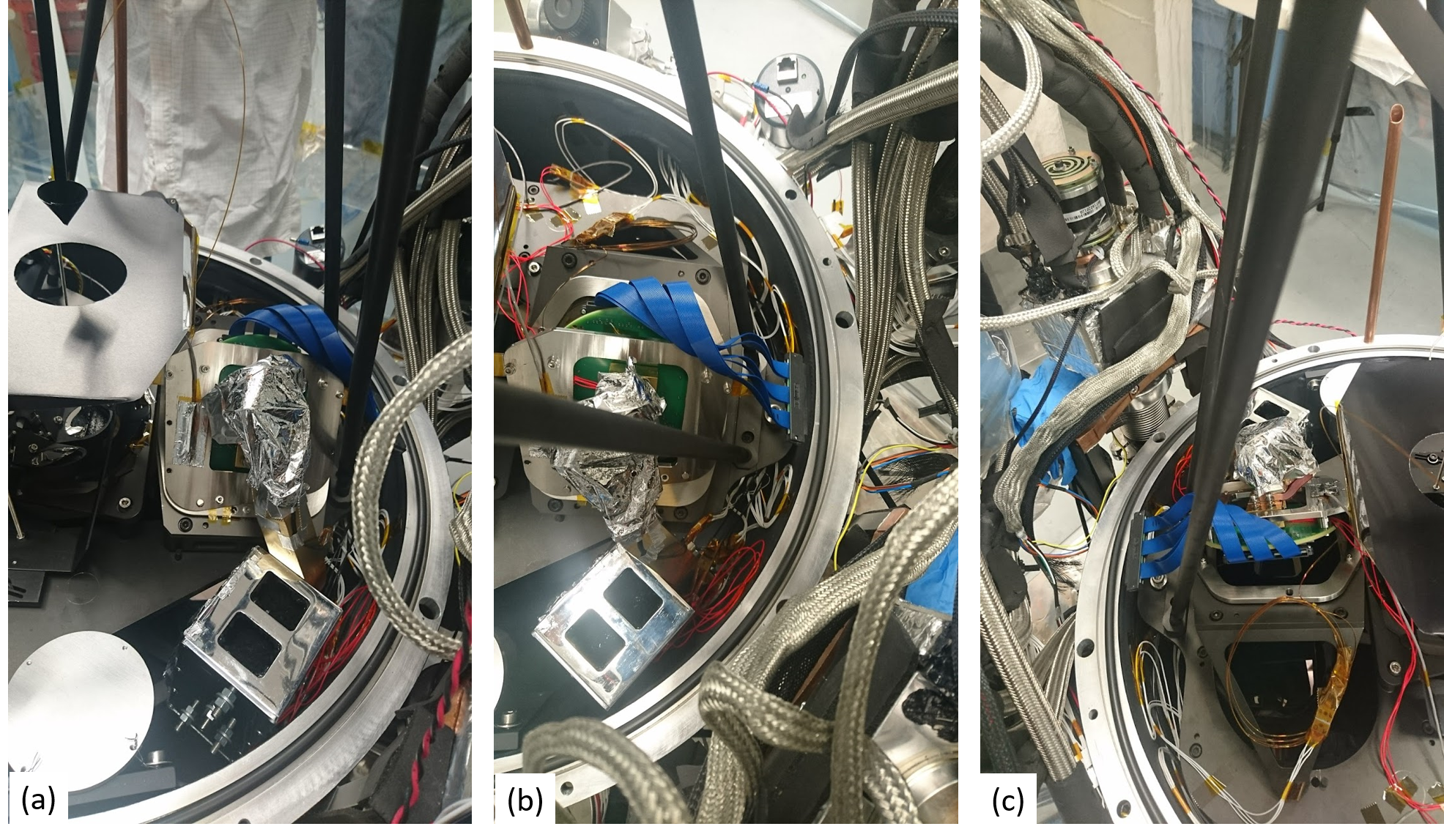} 
\caption{Pictured are a number of different viewpoints of the FB-2 spectrograph tank to show the position and setup of charcoal getter, cold-chain, and detector board and cold-clamp. (a) shows a charcoal getter at the bottom of the image with open windows to provide surface area contact with the vacuum. Also pictured is the cold-chain with two contact points, one to the getter and one to the back of the device. Mylar is used to provide insulation to the cold-chain. (b) shows a different angle to show where the sensors and heaters (placed on the getter and device cold-clamp) connected. The EMCCD PCB pictured in green and SAMTEC cable in blue. (c) from the opposite side of the spectrograph tank. The device cold-clamp is visible and its contact point on the PCB. The SAMTEC cable in blue is pictured and its connection point on the PCB. This cable length is mechanically constrained by the exit point on spectrograph tank to reach the \nuvu\, controller.}
\label{fig:spectro_tank}
\end{figure}

The device is secured to both sides of the PCB. The device is fitted on one side with a gold-plated copper cold-clamp and cooled down to operating temperatures with a CryoTel MT mechanical cryocooler (Sterling Engine) from Sunpower (The CryoTel CT model is used on the FIREBall-2 spectrograph. This is a higher power cooler, which was required for the additional load in flight). A thermal strap feeds from the back of the cold-clamp to cyrocooler coldhead and a silicon diode measures device temperature at the back of the cold-clamp. The thermal strap has a contact with a charcoal getter, which is used as a secondary pumping source in to the main vacuum pumping system (rough pump and turbo pump system). Details of pressure and temperature monitoring achieved can be found in Kyne \cite{Kyne2016}. Figure \ref{fig:spectro_tank} describes the thermal chain, including the interface with the detector.
\par
A Lakeshore unit was used as part of the flight hardware to read out temperature along the cold chain during testing on the ground, though it was not for flight. A PT-100 Resistance Temperature Detector (RTD) sensor was placed on the detector cold-clamp and was calibrated against the Lakeshore diodes, showing no significant differences between the two types of diode sensors. A Ruggeduino microcontroller (based on the Arduino Uno) measures and controls the device temperature both in the laboratory and in FIREBall-2. This measurement is carried out using an RTD shield compatible with the Ruggeduino. Maintaining a stable device temperature is essential when photon counting, as the HV clock is extremely sensitive to temperature changes\cite{Daigle2009}. A solid state relay is controlled by the Ruggeduino to hold the detector temperature at its setpoint throughout data acquisition.

\section{FB-2 detector performance characterization}
This section describes our testing of the FIREBall-2 flight detector from laboratory to flight, including the data reduction methods used. The main goal of testing was to produce a clocking configuration that would achieve the lowest noise profile by minimizing the read noise, dark current, and CIC. We have made improvements to our readout electronics to further reduce our read noise. We have reduced CIC with lower parallel and serial wells, as well as optimizing the clock shaping. Dark current is reduced by cooling, including some work to investigate our clocking configuration as a function of voltage substrate. Testing was performed using the \nuvu V2 controller and will be continued as part of future work using the V3 controller to explore additional HV clocking speeds. \nuvu V3 also has the added advantage of finer clock resolution control, which could help further reduce CIC and dark current.
\par
Noise minimization for the T-e2V CCD201-20 has been investigated previously\cite{Kyne2016}. A number of engineering grade CCD201-20 devices were used to investigate a broad range of clocking options and test the \nuvu V2 controller and a number of PCBs, including the FB-2 flight board. When we had produced a sufficient set of clocking readout sequences using the engineering-grade device, we began noise characterization on a number of delta-doped devices, bare and with and without AR-coatings. The CCD201-20 devices, for back illumination and UV optimization, were processed at JPL; device packaging was split between JPL and T-e2V. The devices were then installed on the PCBs, pumped to a sufficient vacuum (10$^{-6}$ Torr), cooled, and we explored how well certain noise sources, like dark current and CIC, could be controlled with other variables of the system (controller settings, device temperature, etc). We describe the results of our investigation below.

\subsection{CIC}
\label{cic-section}

CIC, or spurious noise, is generated when a CCD is clocked into inversion. Holes become trapped at the Si-SiO$_{2}$ interface and when the clock switches to a noninverted state, the holes are accelerated from the Si-SiO$_{2}$ interface, colliding with the silicon atoms; therefore, CIC is directly related to pixel clocking\cite{Janesick2001}. In a normal CCD, CIC is not apparent because read noise dominates the noise budget ($\sim$ a few electrons). In EMCCDs, CIC is also caused by impact ionization and, therefore, may be counted as an event - indistinguishable from a photoelectron. This is why it is important to minimize CIC: it could lead to false positive detections of what may be interpreted as photon events. Several methods can help reduce CIC, including:

\begin{itemize}
\item Faster clocking speeds. Charge is transferred fast enough, which is less opportunity for a CIC event to be generated.
\item Substrate voltage: More charge carriers are generated in each of the clocks as a result of inversion, which collects any thermal electrons produced.
\item Percentage of clock overlap: The more overlap (50\%) between clock transitions the smoother the charge is transferred.
\item Clock well voltage amplitude: Lower clock amplitude decreases the impact the potential well has on the transfer, thus a smoother transition.
\end{itemize}

Our work to minimize noise contributions from CIC was done in two stages. First, we followed the waveform scripts provided with the \nuvu V2 controller, which used trapezoidal/square waves and a voltage substrate of 4.5V. We found that the parallel CIC dropped by a factor of 10 just by switching to NIMO and using a voltage substrate of 0V.  
\par
Second, we investigated the impact of voltage clock swing for both parallel and serial clocks. Initially, our image clock swings were 12V and serial clocks were 14V. By reducing slowly, we identified a compromise between CIC levels and CTE. A final clock swing of 10 - 11V for each image clock and 12V for the serial clocks resulted in a reduction of CIC $\times$10, and a further $\times$10 reduction when we switched to a sinusoidal clock shape for the image clocks, based on similar successes\cite{Daigle2010}. \nuvu uses an arbitrary clock shape for the serial clocks, and we found this acceptable for our work. Details of this work are given in Kyne \cite{Kyne2016}.

\subsection{Dark Current}

Thermally-generated electrons or dark noise is present in all semiconductor detectors, including conventional CCDs and EMCCDs. Dark current can be reduced by cooling the device during operation and in cooled conventional CCDs, this dark noise is masked by read noise in the same way as CIC. In this way, dark noise will be amplified in EMCCDs, leading to the ability to measure dark rate in photon counting. A significant part of our detector noise testing has focused on measuring CIC in both IMO and NIMO. We find that CIC is minimized in NIMO (see Section~\ref{cic-section}), while dark current is reduced in IMO. For FIREBall-2, we determine the best detector configuration is to operate in NIMO, since dark current can be addressed by further cooling the device. %We have shown (see Section \ref{cic-section}) that it's possible to achieve far lower CIC in NIMO.
\par
To understand the impact of device cooling and dark current rate, we measured the dark current noise level over a range of device temperatures (between -85 and -125$^{\circ}$C). The device temperature is monitored using a silicon diode on the back of the EMCCD cold-clamp, which is read-out with a Pt100 RTD. We found that the temperature differential between device front and back is $\sim$3$^{\circ}$C - this value will vary, depending on the mechanical setups. The temperature differential was calibrated with the same sensor on a non-functional device, for both laboratory and flight hardware.
\par
%The work carried out to optimize dark noise can be found in Kyne \cite{Kyne2016}; with an optimized value for CIC it was possible to begin optimization for dark rate. In brief, 
%The dark current measurements were carried out by taking a series of exposures (5 images) with increasing exposure times \textcolor{red}{(exposure time increase interval? Can you list the number of steps, exposure times, etc?)}. At this point in our work it was also necessary to optimize our data acquisition methods. The software provided with the \nuvu controller worked very well for real time data investigation but it was not optimized for taking multiple data sets continuously. The project also required commandline based software for flight so this was developed and lab tested for data acquisition, and since used for flight.
\par
Each data set for CIC and dark current measurements were acquired using an EMgain of $\sim$1000 e$^{-}$/e$^{-}$. The dark current measurements were carried out by taking a series of exposures (5 images) with increasing exposure times from 0 to 300 seconds, though this was adjusted if longer exposures were required at lower temperatures. Each frame was corrected for cosmic rays (CRs) and smearing (Section~\ref{data_redux}) before a 5.5$\sigma$ threshold was applied and averaged for exposure time. Each full image readout included an pre- and post-scan region on either side of the image area. CIC events were measured from a zero second exposure in the dark current set, and this value was removed from subsequent exposure times to measure dark current. Serial CIC was measured from prescan data. The dark current rate was measured using a least-squares line fit to the plot of events versus exposure time; see Figure \ref{fig:lab_darks}. This same acquisition and analysis was carried out for a series of readouts (varying clock swing and image clock shape), voltage substrate, and temperature.
\par
For both engineering grade and delta-doped T-e2V CCD201-20 devices, we find that dark current rate decreases with device temperature until $\sim$ -110$^{\circ}$C. Below this temperature, we reach a plateau where lowering temperature no longer significantly impacts the dark current rate; in some cases, noise from poor charge transfer causes uncertainty in the absolute dark noise. We also explored moving the clocking into inversion (higher voltage substrates in IMO), and this helped reduce dark current further, but was not as significant as our operating temperatures. The temperature plateau at $\sim$ -115$^{\circ}$C is present for both IMO and NIMO. Further investigation of this behaviour is on-going with \nuvu V3. It is possible that our system suffered from a low level light leak and spurious charge from CR hits, which cannot be removed in analysis. The possibility of a light leak is being further investigated; these results are discussed in the next section.

\subsection{Data reduction and analysis}
\label{data_redux}

A data reduction routine, written in \texttt{Python2.7}, was developed to accurately measure the absolute values of CIC and dark current in our initial laboratory test bed. This reduction pipeline:

\begin{enumerate}
    \item Generates a data cube (A data cube is signal[x,y,t] where t is the same exposure for 5 frames. It is not considered conventional but for this work it was easier to generate a cube like this and make statistical measurements) for each exposure;\label{itm:1}. % time;
    \item Produces a histogram for each data set;
    \item Calculates the electron-multiplying gain (EMgain), FWHM-to-read noise ratio, and data cube bias from the histogram; \label{itm:3}.
    \item Removes cosmic rays;
    \item Generates data cubes from cleaned cosmic ray images per exposure time; \label{itm:5}
    \item Desmears the images for events that were subject to deferred charge/reduced CTE;
    \item Creates a final set of data cubes, generated from desmeared images, where EMgain and $\sigma$ are measured once more;
    \item Thresholds each image using $\sigma$, which generates a cutoff in pixel value and defines the nominal 5.5-$\sigma$ cutoff;
    \item Generates a pixel mask, where a pixel is assigned a 1 if it is above the 5.5-$\sigma$ cutoff and 0 if below; and
    \item Measures CIC from a 0 second exposure (serial-CIC plus parallel-CIC) and dark noise, which are plotted for each subsequent exposure (minus CIC) to measure the noise rate.
\end{enumerate}

The final results produced from this reduction is shown in Figure~\ref{fig:lab_darks}, which shows a CIC-subtracted data set, where a noise over time slope is calculated to extract a dark rate for this clocking configuration and hardware setup (includes detector, PCB and cable, and controller).

\subsection{Cosmic Ray Removal}
Reducing this data involved dealing with the complication of CR removal, which are also subject to amplification and serial register overspill (WFIRST have worked with T-e2V in the development of a version of the CCD201-20 that corrects this overspill in the event of a CR). This means that more than one pixel, and quite often entire rows, are eliminated from a single cosmic rate event. Once steps \ref{itm:1} to \ref{itm:3} above are complete CR removal is carried out.
\par
First, the CRs are counted to estimate a rate, see Figure \ref{fig:flight_cr_rate}. This is done by setting a cutoff value above the bias (a few thousand pixel values); all pixels in the image area above this value are flagged and listed highest to lowest. Starting with the first pixel in this list, successive or connected pixels are identified as part of the same CR until this check fails. These pixels are also flagged. The next pixel in the list is checked in the same way. If 5 or more pixels are found in succession this is characterized as a single CR event and flagged as such. This is repeated until all pixels above the threshold are checked and a final count per image per exposure is generated. Every pixel that has been flagged as part of a CR is used to generate a mask for each CR location in the array.
\par
However, this mask is not enough to correct the raw image and replace the pixel with a random value equal to the bias $\pm \sigma$. To completely remove pixels lost to CRs entire rows are also removed/ignored. This is done based on an average number of rows lost and adding half that number of rows to the to total. In some cases, it was also necessary to conduct a visual inspection of the data to ensure pixels containing charge from a CR are not counted later. Once complete, a new set of data cubes for cleaned CR images are generated per exposure time (Step \ref{itm:5}).

\subsubsection{Smear correction}
The next step in reduction is only used when measuring noise or EMgain from the data. In order to achieve the required low CIC we have worked on reducing both image and serial clock amplitudes, in turn this reduces the CTE. Lower CTE can be measured using $^{55}$Fe, but this only measures CTE in the parallel direction. We have employed a method used by Daigle\cite{Daigle2010} to measure serial CTE in photon counting mode, which has proven a successful metric for the choice of clocking configuration. This will be discussed further in Kyne \cite{Kyne:inpress-a}. As the FB-2 detector operates at a lower than usual temperature, there is a further trade-off in CTE. The FB-2 science is such that our wavelength resolution will not affect the imaging of extended sources, however, EMgain and noise values are affected and must be corrected.
\par
The desmearing algorithm uses a median absolute deviation (MAD) method to measure the spread in pixel value for each image (or data cube). This method is used to investigate a smeared set of pixels, where charge from a photoelectron has been deferred from its pixel of origin during transfer into successive pixels; the length of this pixel set is random and dependent on HV clock value and detector temperature. MAD is a more reliable method for this type of statistical variation because it is not affected by extremely high or extremely low values, and non-normality. This method can be used on a single image or for a data cube of the same exposure. Both work well, with some improved statistics over a large sample set, however, increasing the number of images increases processing time. When the smeared pixels have been flagged an additional algorithm is used to correct the smear and return all the charge to the original pixel; the first in the smeared set to be readout. When the EMgain is measured again it is found to be higher than what was measured from the raw data set, thus smearing greatly affects this measurement. Smearing also affects the dark rate and CIC but it can reduce it in some cases and raise it in others. More details on this can be found in Kyne \cite{Kyne:inpress-a}.

\subsubsection{Thresholding}
\begin{figure}%[h]%{0.25\textwidth}
   % \vspace{-0.3cm}
    \centering
    \caption{The laboratory device, known as w7d10, dark current rate measured in the laboratory at Caltech during clocking optimization.}
    \includegraphics[width=0.5\textwidth]{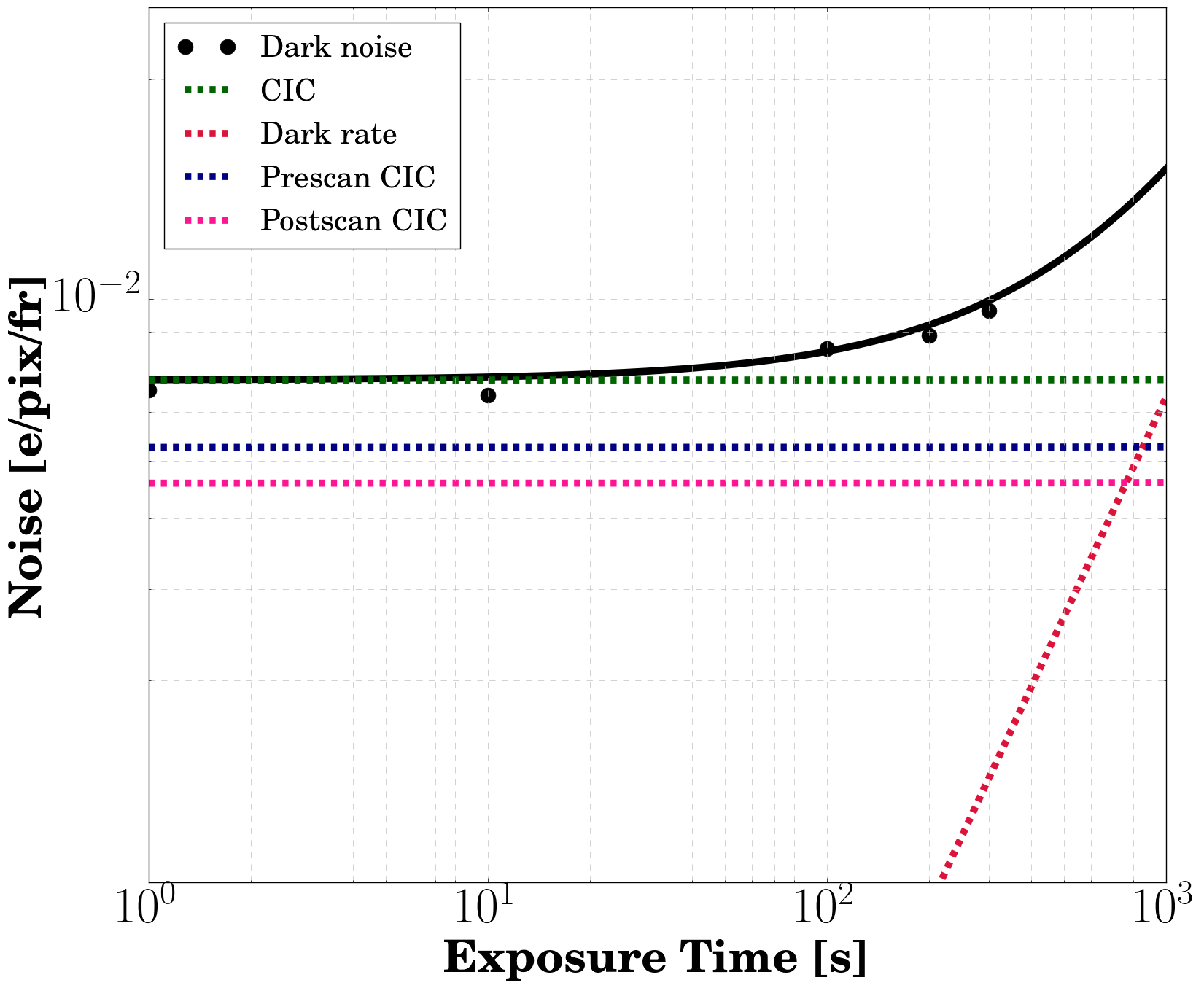}
    \label{fig:lab_darks}
    \vspace{0cm}
\end{figure}

Finally, the data is thresholded, applying a cutoff of chosen $\sigma$ level above the bias and setting any pixel higher than this value to 1; all pixels lower than this value are assigned a 0. An average value for noise is measured for every image, and hence every cube. The CIC is removed and the dark noise is plotted to measure the rate. This is plotted in Figure \ref{fig:lab_darks} and a summary of the noise characterization from our laboratory testing is given in Table \ref{noisetable1} (This table was a baseline for how to clock the detector, including characterization of noise versus VSS level, once an operating temperature was determined).

\begin{table}%[h!]
%\scriptsize{7pt}
\renewcommand{\arraystretch}{1.5}
\caption{Laboratory Noise Characterization: ROS0 (Readout Sequence 0, denotes the readout configuration used to clock the detector for a particular set of parameters) \& Temperature=-110$^{\circ}$C.}
\label{noisetable1}
\begin{center}
\centering
\begin{tabular} {P{2.0cm}||P{2.0cm}||P{2.0cm}||P{2.9cm}||P{2.4cm} }%|P{0.8cm} } %
%\hline
\textbf{VSS [V]} & \textbf{CIC [e$^{-}$/pix/ frame]}  & \textbf{serial CIC [e$^{-}$/pix/ frame]}  & \textbf{Dark Rate [e$^{-}$/pix/hr]} & 
\textbf{Gain [e$^{-}$/e$^{-}$]}  \\ 
\hline\hline
0 & 0.0037 & 0.0029 & 0.025 & 825\\
\hline 
1 & 0.0061 & 0.0054 & 0.017 & 740\\
\hline 
2 & 0.0049 & 0.0042 & 0.011 & 820\\
\hline 
3 & 0.0027 & 0.0023 & 0.018 & 725\\
\hline
4 & 0.0025 & 0.0016 & 0.015 & 745\\
\end{tabular} \\
\end{center}
\end{table}

\section{FIREBall-2: 2018 Flight Campaign}
\label{flight-campaign}

FB-2 launched on a 40 MCF balloon the morning of September 22$^{nd}$, 2018 from Fort Sumner, NM. During the ascent phase (from the surface to float altitude of 128,000 ft), all payload systems and communications behaved nominally. As is normal, the balloon loses altitude as the sun sets (due to changing thermal/pressure conditions), but generally stabilizes at a lower altitude after sunset. For FB-2, which achieves a float altitude during the day of 128 kft, the resulting nighttime float altitude was expected to be around 118 kft. However, it was observed that FB-2 started losing altitude before the sunset transition. Ballast was dropped during sunset to minimize the altitude loss, and the resulting nighttime altitude fell from 118 kft to 105 kft over the course of one hour. The atmospheric transmission in the FB-2 bandpass is highly dependent on altitude, with a reduction of 10\% in transmission for every 9800 ft of altitude lost. In total, FB-2 observed $<$1 hour on science targets before the atmospheric window closed due to altitude loss. Ballast drops did not slow the descent of the payload and created additional instabilities and pointing challenges. While the main cause of the loss in altitude for FB-2 is still under investigation, the most probable culprit appears to be a hole that developed in the balloon sometime after float altitude was achieved during the day.
\par
A hole in the balloon hole also caused a deshape in the balloon profile, from the expected spherical profile to an inverse teardrop shape (which was observed from photos taken from the ground during the flight). Given the availability of launch windows during the narrow, $\sim$2-week long ballooning window out of Ft. Sumner, NM in September, FB-2 also flew during a full moon to take advantage of near-ideal flight wind patterns. Unfortunately, the shape of the deflating balloon, coupled with the full moon, directed moonlight directly past the telescope and into the spectrograph tank. This light entered the spectrograph along an off-axis, direct path to metal (reflective) parts of the detector mount. This light path was not discovered beforehand and, therefore, was not baffled. FB-2 saw an elevated background, due to stray moonlight through the spectrograph, during its entire science operation. A significant post-flight effort is on-going to recreate the observed scattered light profiles seen during flight and to develop new baffling designs to eliminate these stray light paths for future flights. A full discussion of the FB-2 2018 mission and aftermath can be found in Hoadley\cite{Hoadley2019}.

\subsection{Detector Ground Calibrations}

Prior to flight, a large number of calibrations using the UV detector and guider camera were required for geometric mapping. In addition, the EMCCD photon-counting capability required ground calibrations to select the optimal EMgain for flight once it was fully integrated in the spectrograph tank and operating in a flight-like configuration.
\par
A photon transfer curve (PTC) is used to measure the detector conversion gain (electrons per pixel value, in DN or ADU) and read noise (in electrons). EMCCDs, including the CCD201-20, are known to have higher read noise than standard CCDs, where read noise levels in standard CCDs can be as low as 2 electrons. The main contribution to this higher read noise is the higher clocking speed used for EMCCDs (up to 10 MHz) than for standard CCDs ($\sim$100 kHz). For the commercial T-e2V CCD201-20, the design of the readout amplifier in the high gain register is such that the read noise is high -- nearly 100 electrons RMS. The FB-2 PTC was measured using a deuterium hollow-cathode lamp, which covers the entire FB-2 NUV bandpass. The detector conversion gain is dependent on wavelength below 200 nm, so the PTC must be measured carefully with a suitable light source.
\par
The main advantage of using an EMCCD on FB-2 is its photon-counting capability. We achieve this by setting choosing count rate for how many electrons will be produced by the HV clock per photoelectron. We calibrate the HV clock by stepping through a number of clock values and creating a histogram from the image array, which measures the EMgain against the voltage value. This relationship depends on the clocking configuration, making this one of the last detector properties determined before flight operations. 
\par
The choice of EMgain used for flight also depends on the detector read noise. We followed the procedures outlined by Daigle \cite{Daigle2008} (see Figure 1) to settle on a value for $\frac{G}{RN}$. Depending on the chosen threshold cutoff ($\lesssim$ 5.5$\sigma$), $\frac{G}{RN}$ will affect how many events will be counted above this threshold, compared to events that are missed.
\par
\begin{figure}%[h]%{0.25\textwidth}
   % \vspace{-0.3cm}
    \centering
    \caption{Flight device, w17d13, dark current rate measured on the ground in Fort Sumner, NM, one month before the FB-2 flight. The same temperature and clocking configuration as that in Figure \ref{fig:lab_darks} is used for this data set. The measured rate on the ground at this time was 0.13 e$^{-}$/hr.}
    \includegraphics[width=0.5\textwidth]{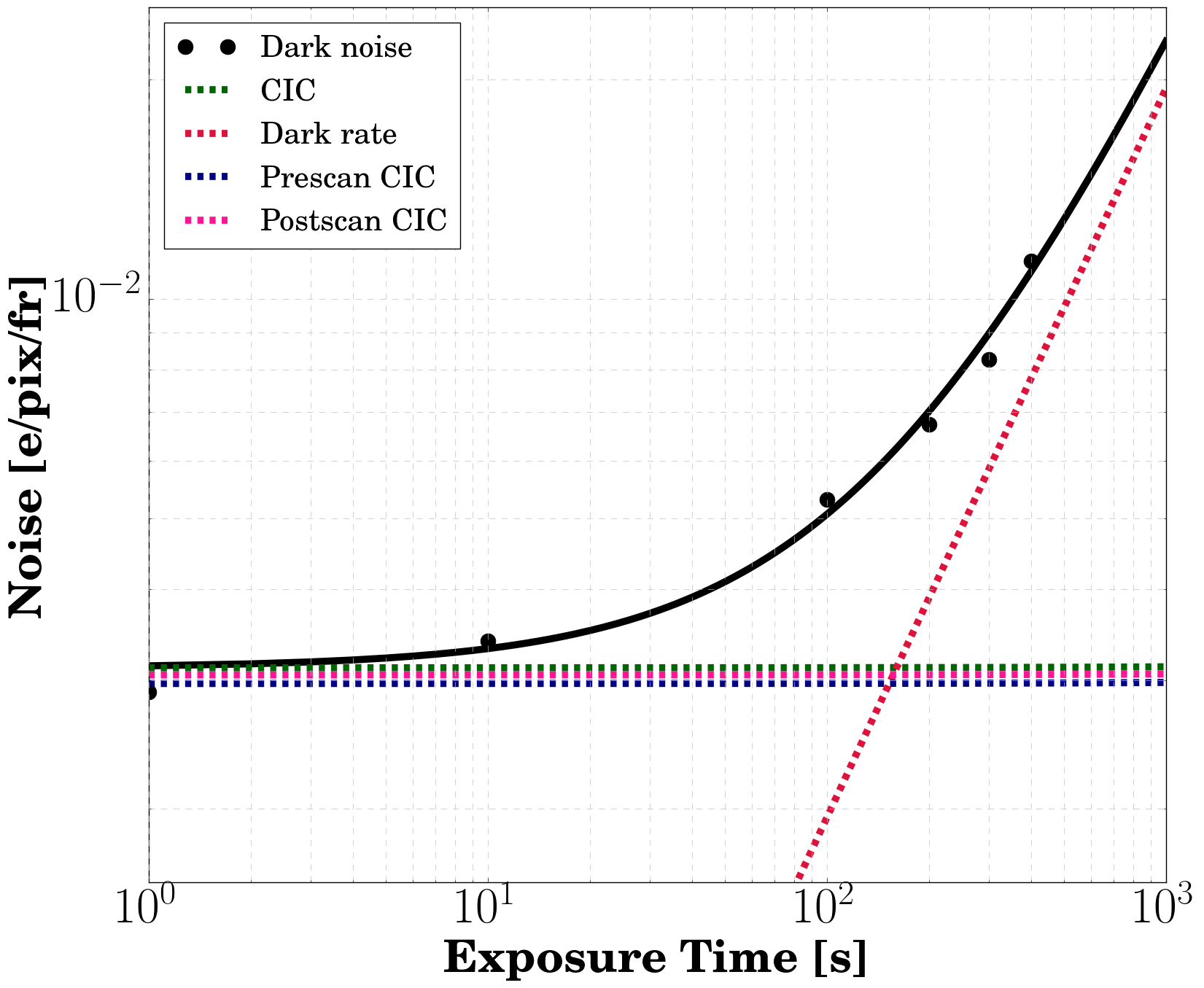}
    \label{fig:gnd_darks}
    \vspace{0cm}
\end{figure}

A number of dark current data sets were obtained in the FB-2 flight configuration to measure CIC and dark rate in preparation for the FB-2 flight. The dark rate plot from these ground dark calibrations is shown in Figure~\ref{fig:gnd_darks}. When compared to Figure~\ref{fig:lab_darks}, we note that the measured dark rate was considerably higher than that measured in the laboratory. We suspect that, at the time the flight-configuration dark current data were taken, the environmental conditions did not allow for entirely light-proof measurements. EMCCDs are very sensitive to low luminosity levels, and any excess light will be amplified in the same way as the dark noise. In addition, the FB-2 CCD201-20 device was selected for its high QE. However, unlike the CCD201-20 device used for laboratory characterization, the flight device has more cosmetics (mostly hot pixels) that likely contribute to an elevated dark rate. At this point, both potential causes for the higher dark rate have not been explored separately, but analysis is on-going.

\subsection{Flight data}

Due to the issues with flight and the defective balloon, FB-2 was unable to achieve its primary mission objectives. However, several important results were achieved including the fact that the whole instrument system, and specifically the delta-doped detector, worked as expected. Significant progress has been made since flight on our understanding of the stray light issues faced by FB-2, as well as astronomical detection limits made by the instrument, both of which are discussed in great detail in Picouet\cite{Picouet2018}. We present some of these results and explain the on-going work below.

\begin{figure}[h!]
\includegraphics[width=1.0\linewidth]{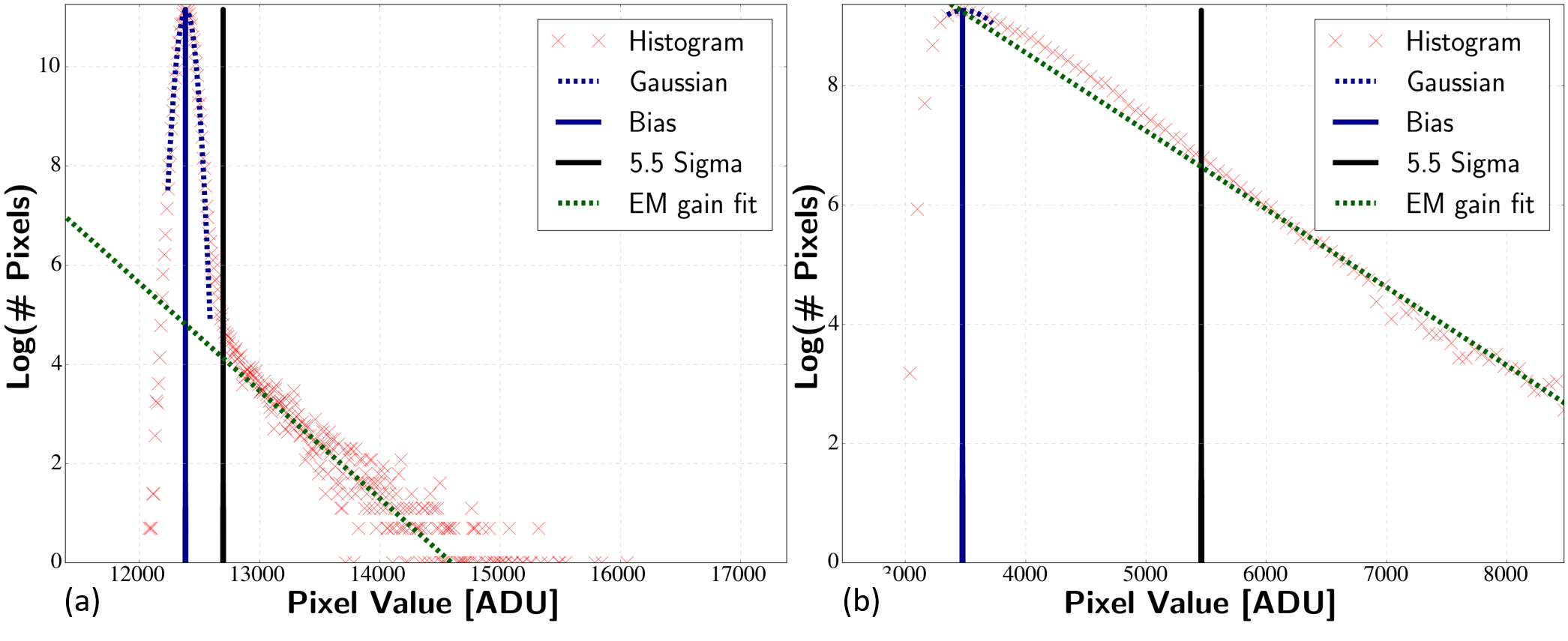} 
\caption{Figure \ref{fig:histograms} (a) shows a desmeared histogram plot of a data cube taken for dark current measurements in the laboratory. This data was acquired at a temperature of -105$^{\circ}$C using the same clocking configuration that would later be used for flight in photon counting mode. However, Figure \ref{fig:histograms} (b) shows that there is a large excess of charge per pixel that dramatically changes the shape of the histogram. This makes EMgain measurement more challenging, however, a measurement for this value from pre-scan data is a useful check; using CIC. We will discuss background subtraction and how this light affects flight data but more detail can be found in Picouet\cite{Picouet2018}.}
\label{fig:histograms}
\end{figure}

\begin{figure}[h!]
\centering
\includegraphics[width=0.85\linewidth]{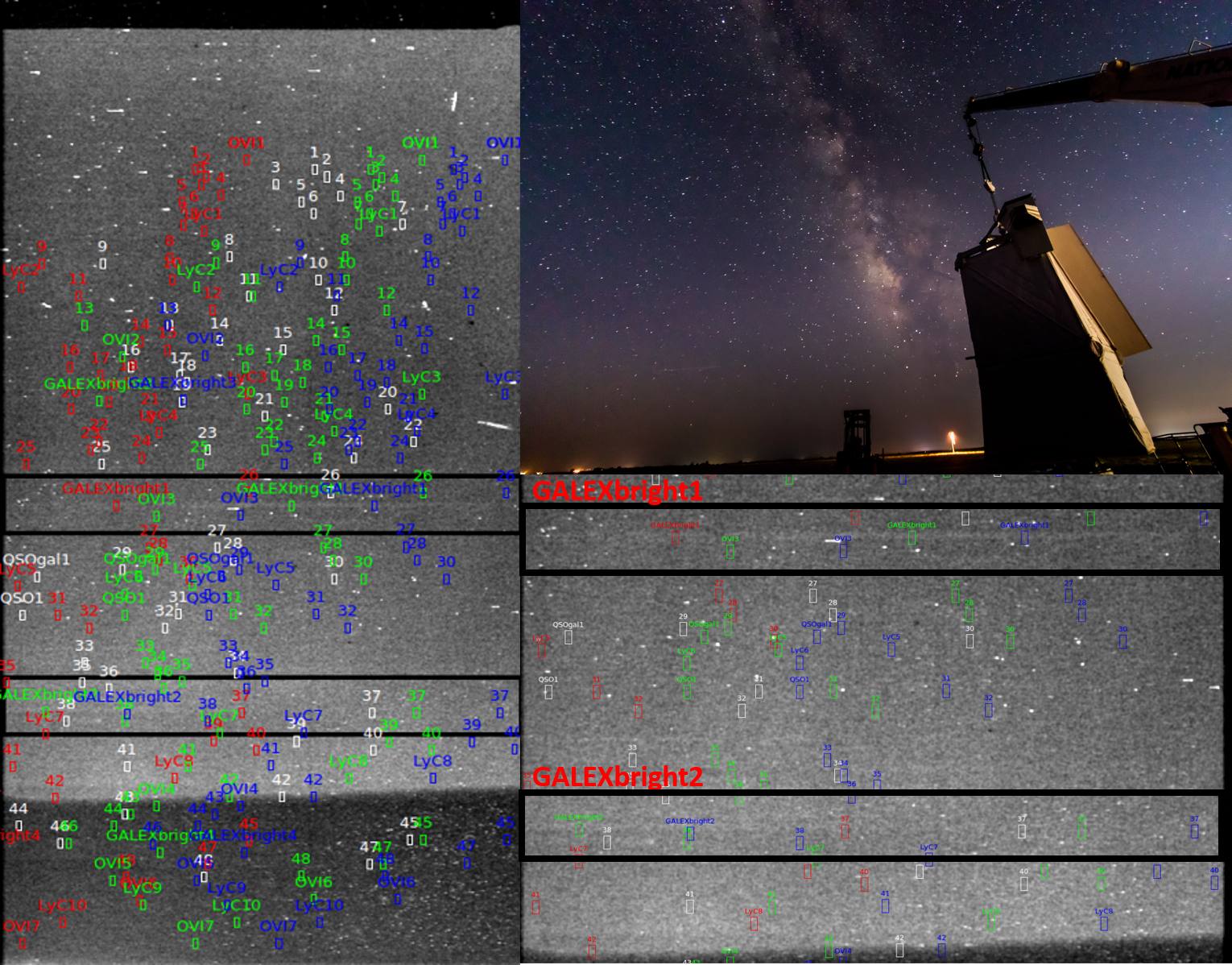} 
\caption{Images of a set of summed data from the 2018 flight campaign. This data is $\sim$ 50 minutes wall clock time. The colored regions have been loaded in \texttt{ds9} to show the locations of absorption/emission lines. The LHS image shows the full FITS file stacked exposure and the RHS bottom image is a zoomed in of the black box regions from this FITS file. Objects GALEXbright 1 and 2 are visible in each of these regions and labelled in red for clarity. These are faint but with improved background subtraction the SNR can be improved. The RHS top image is one of the FB-2 gondola during a hanging sky-test in Fort Sumner, NM with the Milky Way in the background. Image credit: P. Balard.}
\label{fig:summed_data}
\end{figure}

While FB-2 operated its flight device in EMgain mode for photon-counting capabilities, the elevated background level, due to stray moonlight, prevented the detector from operating in photon-counting mode. This is because EMCCDs rely on the statistical detection of photoelectrons over a given exposure, such that the count rate is $<$ 1 e/pix/frame.  Figure \ref{fig:histograms} shows how a histogram describes the problem with a plot of data acquired in the laboratory versus that acquired in flight. With the elevated background levels, the FB-2 EMCCD was seeing $\sim$a few e/pix/frame. Since FB-2 was no longer operating in photon-counting mode, the data from flight could not be thresholded in the normal manner (We note that part of the on-going analysis is investigating whether a modified thresholding method can be used in this event). The current state of the data requires a rigorous background subtraction before science target spectra can be identified and extracted. An initial background subtraction, using a polynomial fit in the y direction (along each column), has been sufficient to reveal some of the brighter calibration targets in the FB-2 science field observed. Picouet\cite{Picouet2018} has demonstrated a more complete subtraction using the Source Extractor Wrapper for Python (\texttt{sewpy}). Figure \ref{fig:summed_data} describes a preliminary data reduction showing a high background but some continuum sources are visible; improvement to this reduction is on-going.

Additionally, there has been a lot of work done to measure the EMgain from flight. The EMgain value is required to convert any given pixel value to electrons. In general, for photon counting data, we can measure the EMgain from the slope of the histogram in the higher signal part of the plot. This mode of operation for an EMCCD is called analog mode: See Figure \ref{fig:emccds_noise_sources} for an explanation on the excess noise factor as a result of operating in this mode. For this process, the dark current measurements must be corrected for smearing in the horizontal register or the inferred EMgain will be underestimated. In the flight data, both smearing effects and excess electrons that are not signal electrons are present and must be accounted for. Post-flight calibration data have been taken at the same temperature and controller settings as the flight data to accurately measure the EMgain in these settings and compare with flight data without excess light. 

\subsection{Cosmic Ray Rates: from laboratory to 110,000 ft}

\begin{figure}[hb]%{0.25\textwidth}
   % \vspace{-0.3cm}
    \centering
    \caption{Plotted is the flight detector cosmic ray rate measured on the ground during the 2018 balloon campaign in Fort Sumner, NM. Not plotted, but based on a large data set of 30 and 50 second exposures, we measured a cosmic ray rate of 5 - 7 cosmic rays per second during flight.}
    \includegraphics[width=0.6\textwidth]{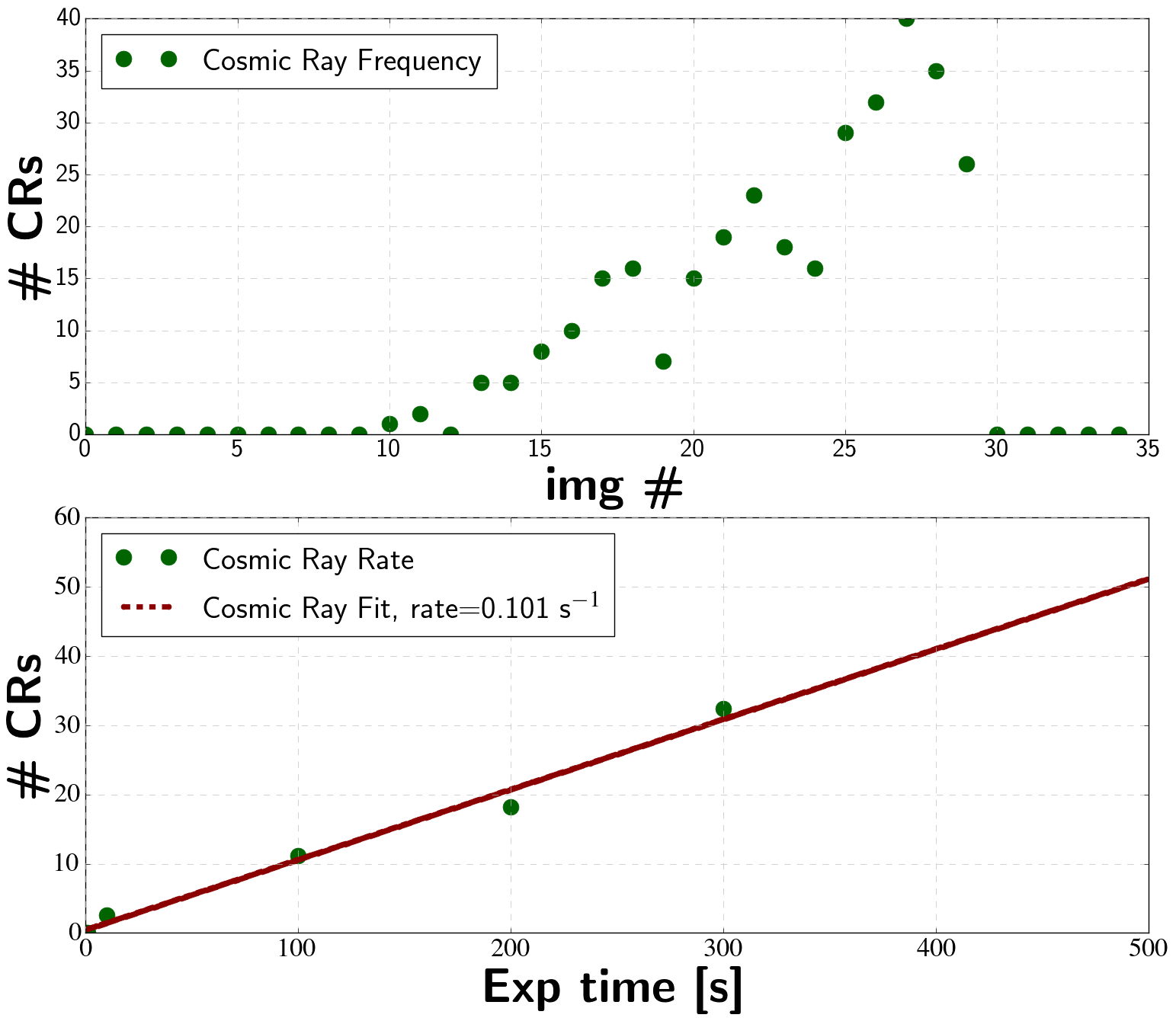}
    \label{fig:flight_cr_rate}
    \vspace{0cm}
\end{figure}

One of the most common sources of noise and loss of image area for CCD exposures is that from CR hits. As has been discussed, the effect of CRs is amplified with the use of HV on EMCCDs. Figure~\ref{fig:flight_cr_rate} shows a plot of CRs on the ground in Fort Sumner, NM during the 2018 campaign. This gave us a baseline measurement before flight where we know to expect an increase in exposure to CRs. There is a separate (ongoing) investigation into the energy distributions of the CRs detected during flight, to understand and implement ways to improve their removal from the images without sacrificing pixels. Here, we present the CR rates measured through different phases of calibration and data collection using the FB-2 EMCCD: During laboratory calibrations at Caltech (elevation: 850 ft above sea level), a cosmic rate of 0.106 s$^{-1}$ was measured. During flight-configuration tests in Ft Sumner, NM (elevation: 4,000 ft above sea level), we measured a CR rate of 0.164 s$^{-1}$. Finally, at float altitudes during the 2018 FB-2 flight (anywhere between 100,000 - 125,000 ft above sea level, as the float altitude did not remain stable), we measured a varying CR rate around 5 -- 7 s$^{-1}$. 
%\par
%We have also discussed future changes in hardware that would mean the CRs would not impact the spectral direction by rotating the device. 

\section{FIREBall-2020}

\subsection{Radiation Testing of 2D delta doped EMCCDs}

Since early 2019, we have been working with the \nuvu V3 controller in preparation for the next flight. Testing of the V3 controller began by using the same clocking configuration to provide a baseline comparison to the FB-2 V2 flight configuration for noise characteristics of the same T-e2V CCD201-20 devices.
\par
The T-e2V delta-doped CCD201-20s are also being prepared for critical radiation tests, slated to happen before the end of calendar year 2019. Proving the performance of these delta-doped EMCCDs in a space-like radiation environment will increase their TRL, paving the way for future space-flight applications. Radiation damage in silicon devices results in traps, which can be mapped and their effect minimized with controlled clocking. A technique called pocket-, or trap-pumping can be used as a way to identify traps, following the work in Murray\cite{Murray2012} and Bush\cite{Bush2018}. We have successfully implemented pocket-pumping using the \nuvu V3 controller with 2 delta-doped devices installed in the same vacuum system. Testing is on-going.
%In addition to characterizing the V3 controllers and boards we are also preparing for radiation testing of these devices for Technology Readiness Level (TRL) advancement of our group's delta-doped EMCCDs. One of our primary goals is to develop these devices for future space missions. This means they must be tested in a radiation like environment, for the extent of a mission. 

\subsection{Goals for next flight}

While the \nuvu V2 controller was proven during the 2018 flight, the next generation \nuvu V3 controller is available with some useful upgrades. Unlike the \nuvu V2 controller, the V3 comes with an option for slower clocking/readout speeds (down to 1 MHz), allowing for lower read noise. We are currently testing different clocking speeds at JPL. The device-to-controller board design is custom-made, such that the same board for both clocking speeds (10 MHz and 1 MHz), is possible. %and using a jumper to connect one at a time, depending on the desired clock speed.
\par
We are currently making improvements to the PCB design for the V3 \nuvu controller including some minor adjustments to the trace thickness and density on the video lines, the distances to ground planes, and the video lines have been to their own plane. We will fabricate these boards and test in our laboratory setup prior to instrument integration, along with the \nuvu V3 controller.
\par
Finally, JPL has fabricated a number of delta-doped CCD201-20 devices with AR coatings optimized for FB-2, and we are investigating  any potential improvement over the last flight device, while still demonstrating the desired ($>$50\%) QE. Our flight device selection was in part dependent on the broader profile of a 3-layer AR coating, however, based on our recent flight results, and calibration data in the field, it is possible to switch to a 5-layer AR coating without major loss in QE in our target bandpass \cite{Jewell2015}.

\section{FIREBall-2 as a pathfinder for future UV telescopes}

The \nuvu V3 CCCP controller, currently being prepared for integration into FB-2 for future flights, has been specified for development and testing as a part of the WFIRST Flagship-class NASA mission for the 2020s. In addition, HabEx, one of four Flagship mission concepts presented for the Astro2020 Decadal review, has the same EMCCD technology and \nuvu controller specified for their UV spectrograph. FB-2 provides a key platform to test, demonstrate, and mature NASA space mission technologies, according to the Cosmic Origins Annual Technology Report\cite{NASA2017}.

\acknowledgments{The research was carried out at the Jet Propulsion Laboratory, California Institute of Technology, under a contract with the National Aeronautics and Space Administration. The authors would like to acknowledge generous and excellent collaborative support from e2v including A. Reinheimer, P. Jerram, P. Jorden, and their team on the 2-D doped EMCCDs.}

%The research was carried out in part at the Jet Propulsion Laboratory, California Institute of Technology, under a contract with the National Aeronautics and Space Administration.}

\bibliography{longtitles,emccd_bibliography}
\bibliographystyle{spiejour}   % makes bibtex use spiejour.bst

%%%%% Biographies of authors %%%%%

\vspace{2ex}\noindent\textbf{Gillian Kyne} is the detector scientist for the FIREBall-2 scientific balloon telescope. She received her BS \textbf{[in physics]} from \textbf{[The National University of Ireland, Galway]} in \textbf{[2009]} and her PhD degree \textbf{[in astronomy]} from the same university in \textbf{[2014]}. She is an expert in detector characterization and building astronomical instruments. Her research interests include \textbf{[designing and building hardware for the laboratory, developing new ways and techniques to make current hardware more efficient. In addition, writing software control systems and developing data reduction and analysis code]}.

\vspace{1ex}
\noindent Biographies and photographs of the other authors are not available.

\listoffigures
\listoftables
\end{spacing}
\end{document}